\documentclass[pra,twocolumn,superscriptaddress]{revtex4}
\usepackage{amsfonts}
\usepackage{amssymb}
\usepackage{amsmath}
\usepackage{epsfig}
\usepackage{color}
\usepackage{graphics, graphicx}
\usepackage{bbold}
\usepackage{psfrag}
\usepackage{mathcomp}
\usepackage{subfigure}
\usepackage{verbatim}
\usepackage[colorlinks,citecolor=blue]{hyperref}

\makeatletter

\newcommand{\Rmnum}[1]{\expandafter\@slowromancap\romannumeral #1@}
\makeatother

\begin{document}

\title{Topological nodal chains in optical lattices}

\author{Bei-Bei Wang}

\affiliation{State Key Laboratory of Quantum Optics and Quantum Optics Devices, Institute
of Laser Spectroscopy, Shanxi University, Taiyuan, Shanxi 030006, China}
\affiliation{Collaborative Innovation Center of Extreme Optics, Shanxi
University, Taiyuan, Shanxi 030006, China}

\author{Jia-Hui Zhang}

\affiliation{State Key Laboratory of Quantum Optics and Quantum Optics Devices, Institute
of Laser Spectroscopy, Shanxi University, Taiyuan, Shanxi 030006, China}
\affiliation{Collaborative Innovation Center of Extreme Optics, Shanxi
University, Taiyuan, Shanxi 030006, China}

\author{Chuanjia Shan}
\email{cjshan@hbnu.edu.cn}
\affiliation{College of Physics and Electronic Science, Hubei Normal University, Huangshi 435002, China}

\author{Feng Mei}
\email{meifeng@sxu.edu.cn}
\affiliation{State Key Laboratory of Quantum Optics and Quantum Optics Devices, Institute
of Laser Spectroscopy, Shanxi University, Taiyuan, Shanxi 030006, China}
\affiliation{Collaborative Innovation Center of Extreme Optics, Shanxi
University, Taiyuan, Shanxi 030006, China}

\author{Suotang Jia}
\affiliation{State Key Laboratory of Quantum Optics and Quantum Optics Devices, Institute
of Laser Spectroscopy, Shanxi University, Taiyuan, Shanxi 030006, China}
\affiliation{Collaborative Innovation Center of Extreme Optics, Shanxi
University, Taiyuan, Shanxi 030006, China}

\date{\today}

\begin{abstract}
  Topological nodal rings as the simplest topological nodal lines recently have been extensively studied in optical lattices. However, the realization of complex nodal line structures like nodal chains in this system remains a crucial challenge. Here we propose an experimental scheme to realize and detect topological nodal chains in optical Raman lattices. Specifically, we construct a three-dimensional optical Raman lattice which supports next nearest-neighbor spin-orbit couplings and hosts topological nodal chains in its energy spectra. Interestingly, the realized nodal chains are protected by mirror symmetry and could be tuned into a large variety of shapes, including the inner and outer nodal chains. We also demonstrate that the shapes of the nodal chains could be detected by measuring spin polarizations. Our study opens up the possibility of exploring topological nodal-chain semimetal phases in optical lattices.

\end{abstract}

\maketitle

\section{Introduction}

The study of topological nodal line semimetals have recently emerged as a new frontier in condensed matter physics \cite{TNLrev,TNLBook}. Nodal line semimetals are systems where the conduction and the valence bands touch along closed nodal lines in the Brillouin zone \cite{TNLrev,TNLBook}. Intriguingly, such gapless phase also possesses topological properties, such as quantized Berry phase and topologically protected drumhead surface states \cite{TNL}. Nodal lines have numerous topological structures than nodal points, including nodal rings \cite{NR1,NR2,NR3,Volovik2015,NR4,Volovik2018,Volovik2019,Volovik2020}, nodal chains \cite{NC1,NC2,NC3,NC4,NC5,NC6,NC7,NC8}, nodal links \cite{NL1,NL2,NL3,NL4,NL5} and nodal knots \cite{NK1,NK2}. In particular, the nodal chains are formed by the nodal rings located on mutually orthogonal high-symmetry planes and touching each other at isolated points \cite{NC1,NC2,NC3},
which are the basics of generating nodal links and knots \cite{NC3,NC8}. Thus, nodal chains have much richer topological properties waiting to be explored and observed, such as the unusual topological electromagnetic responses \cite{NC1} and distinct topological surfaces states \cite{NC3,NC6,NC7}. Experimental efforts recently have been devoted to searching materials with different nodal lines \cite{TNLexp1,TNLexp2}. Unfortunately, the nodal lines showing there are often not exactly located at the Fermi level as the ideal one, mixed with the trivial bulk bands, making the experimental detection much challenge.

Meanwhile, ultracold atoms trapped in optical lattices provides a clean and controllable platform for realizing, manipulating and probing topological phases of matter \cite{TPCArev1,TPCArev2,TPCArev3}. This has been well demonstrated by the recent experimental implementation of spin-orbit couplings \cite{SOC1,SOC2,SOC3,SOC4,SOC5,SOC6,SOC7} and topological insulator phases \cite{SSHCA,HaldaneCA,HofstaCA,SOCTI1CA,SOCTI2CA} in low-dimensional ultracold atomic systems. The state-of-the-art optical lattice technology developed in these experiments further paves the way for exploring high-dimensional gapless topological phases \cite{Xu}. In particular, a lot of optical lattice models have been constructed for realizing topological nodal rings \cite{NRCA1,NRCA2,NRCA3,NRCA4,NRCA5,NRCA6,NRCA7}. Since optical lattices naturally have clean environments, the realized nodal rings are exactly located at the Fermi level and not mixed with trivial bulk bands. However, how to realize more exotic nodal line structures like nodal chains in this system is still unknown.

Very recently, two experiments reported the realization and detection of topological nodal lines \cite{NLexp} and Weyl points \cite{WPexp} with ultracold atoms trapped in optical Raman lattices. Motivated by these experiments, we present a three-dimensional optical Raman lattice system that could generate next nearest-neighbor spin-orbit couplings, which has not been reported before. Moreover, the corresponding tight-binding Hamiltonian has not been previously discovered in solid state material or optical lattice systems. Distinct from previous nodal-ring studies in optical lattices \cite{NRCA1,NRCA2,NRCA3,NRCA4,NRCA5,NRCA6,NRCA7}, our system could host a lot of nodal rings which are located on mirror invariant planes and protected by the mirror symmetry. By tuning the laser intensities and frequencies to vary the hopping rates and effective Zeeman field, the nodal rings on the mutually orthogonal mirror invariant planes are connected together to form various shapes of topological nodal chains, including the inner and outer nodal chains. In addition, we also demonstrate that the shapes of the nodal chains could be detected through the measurement of spin polarizations.

The paper is organized as follows. Section \ref{secII} presents an optical Raman lattice model with next nearest-neighbor spin-orbit couplings. Section \ref{secIII} shows the energy spectra hosts mirror symmetry protected topological nodal rings and chains. Section \ref{secIV} discusses the experimental detection of the nodal chains. Section \ref{secV} summarizes the main results of this paper.

\begin{figure}
\centering
\includegraphics[width=8.5cm,height=4cm]{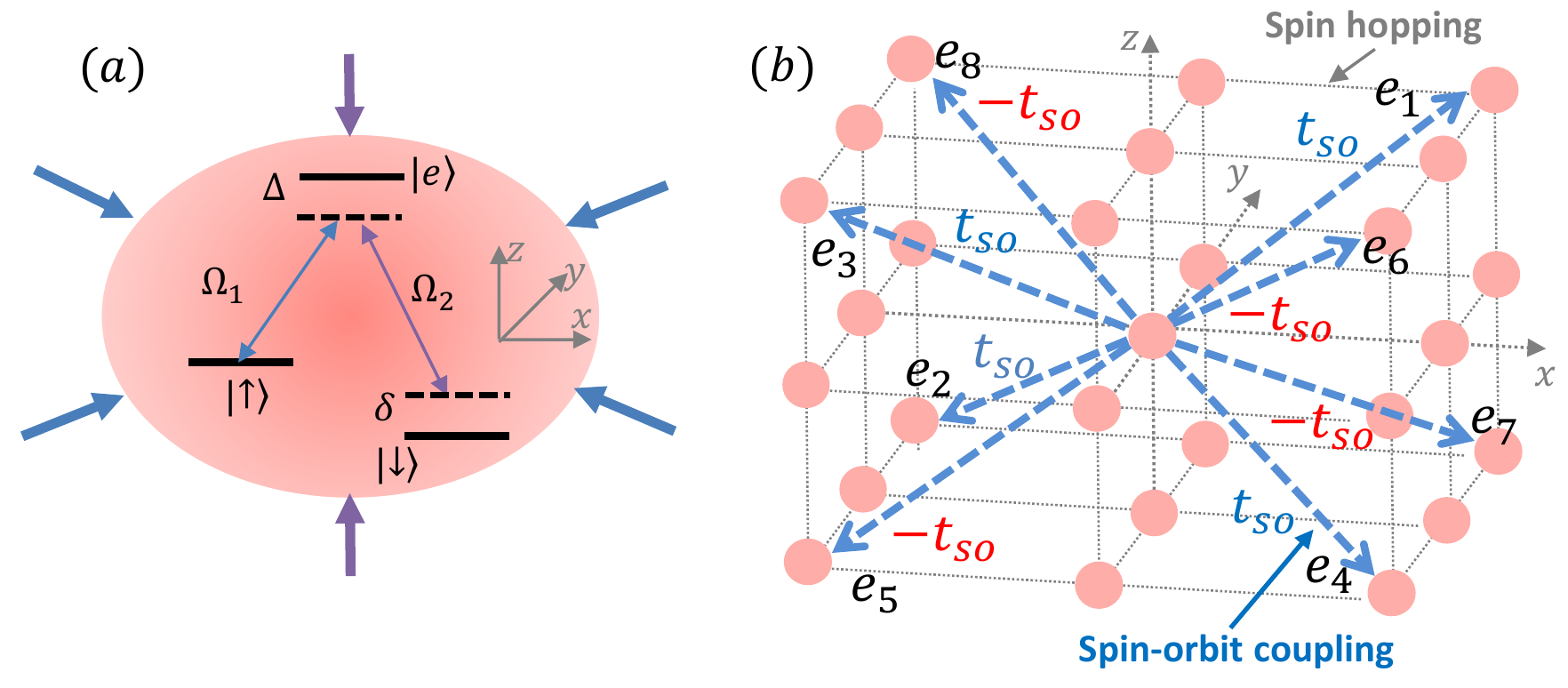}
\caption{(a) Schematics of the laser configuration to realize the two-photon Raman coupling. Such coupling is generated by two laser beams with Rabi frequencies $\Omega_{1}$ and $\Omega_{2}$. $\Delta$ is the detuning from the auxiliary excited state and $\delta$ is the two-photon detuning in the Raman transition. (b) In the cubic optical lattice with the Raman coupling, the spin-orbit couplings are generated only between the lattice sites $\emph{\textbf{j}}$ and $\emph{\textbf{j}}^{\prime}=\emph{\textbf{j}}+\emph{\textbf{e}}_v$ ($v=1,2,3,4,5,6,7,8$). Moreover, the generated synthetic spin-orbit couplings are position-dependent, i.e., the strengthes of the spin-orbit couplings along $\emph{\textbf{e}}_v$ for $v=1,2,3,4$ ($v=5,6,7,8$) are $t_{so}$ ($-t_{so}$).
 }
\label{Fig1}
\end{figure}

\section{Optical Raman lattices with next nearest-neighbor spin-orbit couplings}
\label{secII}

We consider using the experimental setup reported for observing ideal Weyl points \cite{WPexp}. This setup is made up of ultracold atoms trapped in optical Raman lattices. In addition to conventional optical lattice trapping potentials, an effective Raman lattice potential is also applied in optical Raman lattice systems \cite{XJL2013,XJL2014,XJL2018}. Suppose a spin is encoded by two atomic internal states. The effective Raman potential is generated through a two-photon Raman transition which couples the spin up and the spin down \cite{XJL2013,XJL2014,XJL2018}. Here we find that, by changing the Raman potential applied in \cite{WPexp}, the resulted optical Raman lattice could generate next nearest-neighbor spin-orbit couplings and support various topological nodal rings and chains.

The experimental setup is shown in Fig. \ref{Fig1}(a), where ultracold $^{87}$Rb atoms are trapped in a three-dimensional optical Raman lattice. The spin is encoded by two magnetic sublevels $\mid\uparrow\rangle=|1,-1\rangle$ and $\mid\downarrow\rangle=|1,0\rangle$ \cite{WPexp}, with an energy difference $\omega_0$. The conventional cubic optical lattice is generated by applying three standing wave lasers in the three real space directions, leading to a trapping potential (spin-independent) $V_{u=x,y,z}=\bar{V}_{u}\cos^2(k_0u)$, where $\bar{V}_{u}\propto |E_u|^2$ and $E_u$ is the amplitude of the standing wave laser in the $u$ direction. Furthermore, the two-photon Raman transition between the spin up and the spin down is generated by applying two standing wave lasers with frequency $\omega_1$ in the $x$-$y$ plane and one standing wave laser with frequency $\omega_2$ in the $z$ direction. The corresponding Rabi frequencies are $\Omega_{1}=\bar{\Omega}_{1}\cos(k_0x)\cos(k_0y)$ and $\Omega_{2}=\bar{\Omega}_{2}\cos(k_0z)$ respectively, leading to a Raman coupling potential (spin-dependent) $V_{so}\sigma_x$, where $V_{so}=\Omega_{so}\cos(k_0x)\cos(k_0y)\cos(k_0z)$ and $\Omega_{so}=\bar{\Omega}_{1}\bar{\Omega}_{2}/\Delta$.
Then the total Hamiltonian of the system reads
\begin{equation}
H_s=\sum_{u=x,y,z}\big(\frac{p^2_u}{2m}+V_u\big)+V_{so}\sigma_x+m_z\sigma_z,
\label{Hs}
\end{equation}
where $m_z=\delta/2$ is the effective Zeeman field and $\delta=\omega_1+\omega_0-\omega_2$ is the two-photon Raman detuning.

In the second quantization, the continuum Hamiltonian $H_s$ is written into the following form
\begin{equation}
H=\int d\emph{\textbf{r}}\psi^{\dag}(\emph{\textbf{r}})H_s\psi(\emph{\textbf{r}})
\label{Hsq}
\end{equation}
where $\psi(\emph{\textbf{r}})=(\psi_{\uparrow}(\emph{\textbf{r}}),\psi_{\downarrow}(\emph{\textbf{r}}))^T$. Here we only study physics in the ground band. Then the field operator can be expanded as
\begin{equation}
\psi_{\sigma}(\emph{\textbf{r}})=\sum_{\emph{\textbf{j}}}W(\emph{\textbf{r}}-\emph{\textbf{j}})C_{\emph{\textbf{j}}\sigma},
\label{psi}
\end{equation}
where $W(\emph{\textbf{r}}-\emph{\textbf{j}})$ is the spin-independent Wannier function for the ground band located at the lattice site $\emph{\textbf{j}}=(j_x,j_y,j_z)$. Throughout this work, we assume the lattice spacing $a=\pi/k_0=1$. $C_{\emph{\textbf{j}}\sigma}$ is the annihilation operator for an atom with spin $\sigma=\uparrow,\downarrow$ at the lattice site $\emph{\textbf{j}}$. With the above expansion, we can get the tight-binding Hamiltonian
\begin{align}
H=&-\sum_{\emph{\textbf{j}}}\sum_{u=x,y,z}t_u(C^{\dag}_{\emph{\textbf{j}}\uparrow}C_{\emph{\textbf{j}}
+\emph{\textbf{e}}_u\uparrow}+
C^{\dag}_{\emph{\textbf{j}}\downarrow}C_{\emph{\textbf{j}}
+\emph{\textbf{e}}_u\downarrow}+h.c.) \nonumber \\
&+\sum_{\emph{\textbf{j}},\emph{\textbf{j}}^{\prime}}t_{so}^{\emph{\textbf{j}},\emph{\textbf{j}}^{\prime}}
(C^{\dag}_{\emph{\textbf{j}}\uparrow}
C_{\emph{\textbf{j}}^{\prime}\downarrow}+h.c)
+m_z\sum_{\emph{\textbf{j}}}(C^{\dag}_{\emph{\textbf{j}}\uparrow}
C_{\emph{\textbf{j}}\uparrow}-C^{\dag}_{\emph{\textbf{j}}\downarrow}
C_{\emph{\textbf{j}}\downarrow}),
\label{tbh0}
\end{align}
where the vectors $\emph{\textbf{e}}_x=(1,0,0)$, $\emph{\textbf{e}}_y=(0,1,0)$ and $\emph{\textbf{e}}_z=(0,0,1)$, the nearest-neighbor hopping rate and the spin-orbit coupling strength are expressed as
\begin{align}
&t_u=-\int du W_u^*(u)(\frac{p^2_u}{2m}+V_u)W_u(u-1), \\
&t_{so}^{\emph{\textbf{j}},\emph{\textbf{j}}^{\prime}}
=\Omega_{so}\prod_{u=x,y,z}t_{so,u}^{j_u,j_u^{\prime}},
\label{tso}
\end{align}
with
\begin{align}
t_{so,u}^{j_u,j_u^{\prime}}&=\int du\, W_u^*(u-j_u)\cos(k_0u)W_u(u-j_u^{\prime}) \nonumber \\
&=\int du\, W_u^*(u)\cos(k_0(u+j_u))W_u(u+j_u-j_u^{\prime}) \nonumber \\
&=-(-1)^{j_u}\int du\, W_u^*(u)\sin(k_0u)W_u(u+j_u-j_u^{\prime}).
\label{tsojj}
\end{align}
Here we use the identities $W(\emph{\textbf{r}}-\emph{\textbf{j}})=W_x(x-j_x)W_y(y-j_y)W_z(z-j_z)$ and $\cos^2(k_0j_u)=0$. The last identity is obtained because the bottom of the lattice trapping potential is the lattice site. In optical lattices, we usually only consider $j_u-j_u^{\prime}=0, \pm1$, because the integral in Eq. (\ref{tsojj}) for $j_u-j_u^{\prime}>1$ is much smaller than the one for $j_u-j_u^{\prime}=0,\pm1$.

Considering that the Wannier function $W_u$ for the ground band has an even parity, basing on Eq. (\ref{tsojj}), we obtain
\begin{align}
&t_{so,u}^{j_u,j_u}=0,  \label{tsorela1} \\
&t_{so,u}^{j_u,j_u+1}=-t_{so,u}^{j_u,j_u-1}.
\label{tsorela2}
\end{align}
Combining Eq. (\ref{tso}) and Eq. (\ref{tsorela1}), we further get
\begin{align}
&t_{so}^{\emph{\textbf{j}},\emph{\textbf{j}}}=0, \nonumber \\
&t_{so}^{\emph{\textbf{j}},\emph{\textbf{j}}+\emph{\textbf{e}}_x}=
t_{so}^{\emph{\textbf{j}},\emph{\textbf{j}}+\emph{\textbf{e}}_y}=
t_{so}^{\emph{\textbf{j}},\emph{\textbf{j}}+\emph{\textbf{e}}_z}=0, \nonumber \\
&t_{so}^{\emph{\textbf{j}},\emph{\textbf{j}}+\emph{\textbf{e}}_x+\emph{\textbf{e}}_y}=
t_{so}^{\emph{\textbf{j}},\emph{\textbf{j}}+\emph{\textbf{e}}_x+\emph{\textbf{e}}_z}=
t_{so}^{\emph{\textbf{j}},\emph{\textbf{j}}+\emph{\textbf{e}}_y+\emph{\textbf{e}}_z}=0.
\label{tsorela3}
\end{align}
From the above equations, we can see that the spin-orbit couplings are mainly generated between $\emph{\textbf{j}}=(j_x,j_y,j_z)$ and $\emph{\textbf{j}}^{\prime}=(j_x\pm1,j_y\pm1,j_z\pm1)$. Specifically, as shown in Fig. \ref{Fig1}(b), the spin-orbit couplings act only along the vectors $\emph{\textbf{e}}_1=-\emph{\textbf{e}}_5=(1,1,1)$, $\emph{\textbf{e}}_2=-\emph{\textbf{e}}_6=(-1,1,-1)$, $\emph{\textbf{e}}_3=-\emph{\textbf{e}}_7=(-1,-1,1)$ and $\emph{\textbf{e}}_4=-\emph{\textbf{e}}_8=(1,-1,-1)$. Combining Eq. (\ref{tso}) and Eq. (\ref{tsorela2}), we also can find that the generated spin-orbit couplings are position-dependent and have the following relationship
\begin{equation}
t_{so}^{\emph{\textbf{j}},\emph{\textbf{j}}+\emph{\textbf{e}}_s}=
-t_{so}^{\emph{\textbf{j}},\emph{\textbf{j}}+\emph{\textbf{e}}_t},
\label{tsoiden}
\end{equation}
where $s=1,2,3,4$ and $t=5,6,7,8$. By substituting Eqs. (\ref{tso},\ref{tsojj},\ref{tsoiden}) into Eq. (\ref{tbh0}) and performing a gauge transformation $(-1)^{j_x+j_y+j_z}C_{\emph{\textbf{j}}\downarrow}\rightarrow C_{\emph{\textbf{j}}\downarrow}$ \cite{XJL2013,XJL2014}, the tight-binding Hamiltonian becomes
\begin{align}
H=&-\sum_{\emph{\textbf{j}}}\sum_{u=x,y,z}t_u(C^{\dag}_{\emph{\textbf{j}}\uparrow}C_{\emph{\textbf{j}}
+\emph{\textbf{e}}_u\uparrow}-
C^{\dag}_{\emph{\textbf{j}}\downarrow}C_{\emph{\textbf{j}}
+\emph{\textbf{e}}_u\downarrow}+h.c.) \nonumber \\
&+\sum_{\emph{\textbf{j}}}\sum^4_{v=1}t_{so}(C^{\dag}_{\emph{\textbf{j}}\uparrow}
C_{\emph{\textbf{j}}+\emph{\textbf{e}}_v\downarrow}
-C^{\dag}_{\emph{\textbf{j}}\uparrow}
C_{\emph{\textbf{j}}-\emph{\textbf{e}}_v\downarrow}+h.c.)\nonumber \\
&+m_z\sum_{\emph{\textbf{j}}}(C^{\dag}_{\emph{\textbf{j}}\uparrow}
C_{\emph{\textbf{j}}\uparrow}-C^{\dag}_{\emph{\textbf{j}}\downarrow}
C_{\emph{\textbf{j}}\downarrow}),
\label{tbh}
\end{align}
where $t_{so}=\Omega_{so}t_{so,x}^{01}t_{so,y}^{01}t_{so,z}^{01}$. The hopping rates, the spin-orbit coupling strengths and the effective Zeeman fields all can be flexibly tuned by changing laser intensities and frequencies.

In contrast to previous experiments producing nearest-neighbor spin-orbit couplings \cite{SOCTI1CA,SOCTI2CA,NLexp,WPexp}, our proposed optical Raman lattice could generate the next nearest-neighbor spin-orbit couplings between the lattice sites $\emph{\textbf{j}}=(j_x,j_y,j_z)$ and $\emph{\textbf{j}}^{\prime}=(j_x\pm1,j_y\pm1,j_z\pm1)$, which are out of the $x$, $y$ and $z$ planes (see Fig. \ref{Fig1}(b)). The reason is that, the Wannier function $W_{u=x,y,z}$ for the ground band has an even parity and the Raman potential $V_{so}$ centered at the lattice site has an odd parity (see Eq. (\ref{tsojj})), which leads to the vanishing of the integrals in Eq. (\ref{tso}) for the on-site spin flips and the nearest-neighbor spin-orbit couplings (see Eq. (\ref{tsorela3})). Thanks to these particular next nearest-neighbor spin-orbit couplings, as we will demonstrate below, the nodal chains protected by mirror symmetry are generated in this system.

\begin{figure*}
\centering
\includegraphics[width=16cm,height=11cm]{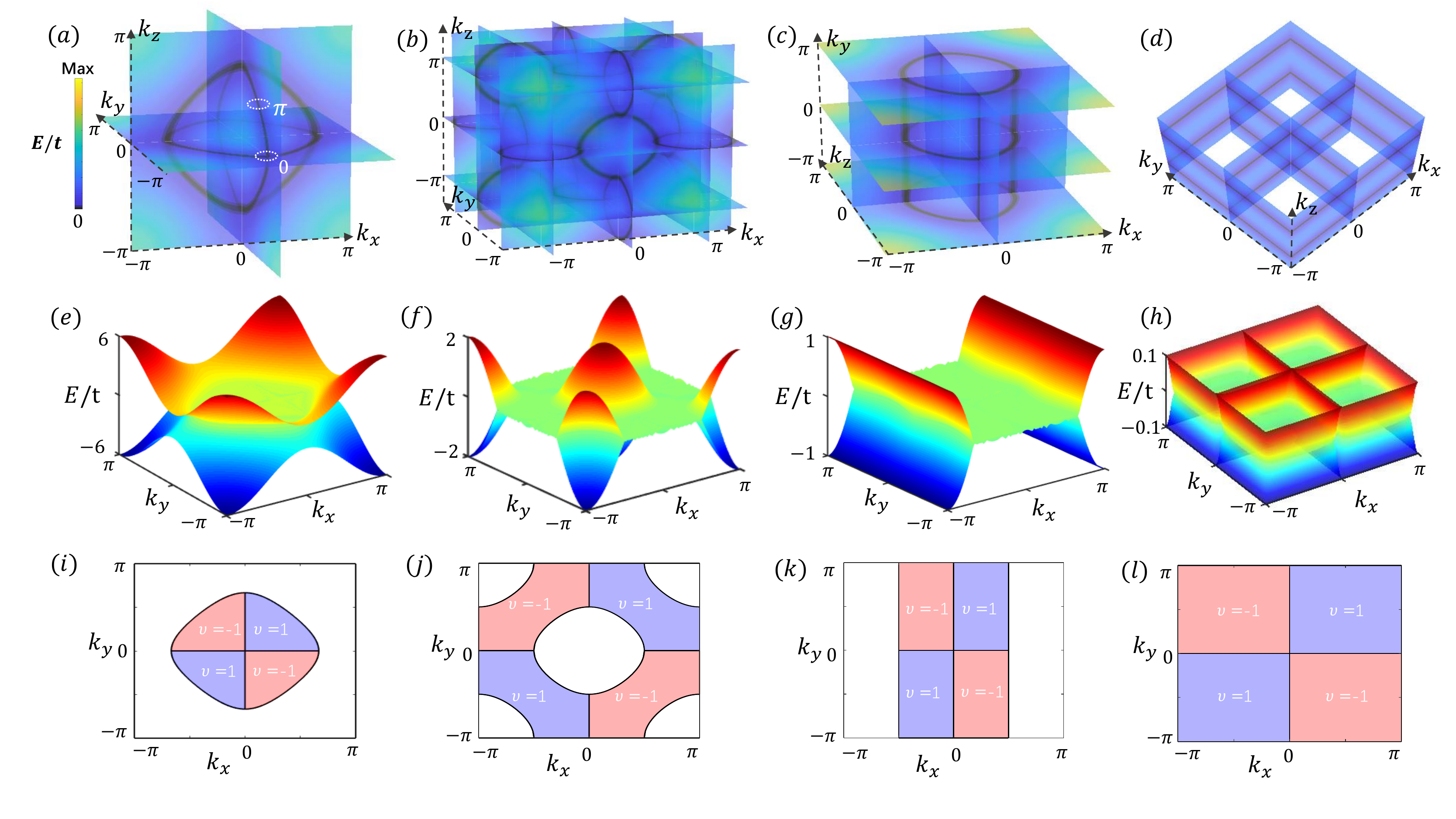}
\caption{The nodal rings and chains in the energy spectra for (a) $t_x=t_y=t$, $m_z=3t$, (b) $t_x=t_y=t$, $m_z=0$, (c) $t_x=t$, $t_y=0$, $m_z=2t$ and (d) $t_x=t_y=m_z=0$. Specifically, the nodal rings (gap-closing points marked by black solid lines) are generated on the mirror invariant planes $k_{x,y,z}=0,\pi$. The topological features of nodal rings are characterized by the quantized Berry phases along a tiny loop enclosing the nodal lines. Our numerical results show that the quantized Berry phase is $\pi$ ($0$) when the loop encloses one (two) nodal lines. By tuning the hopping rates and effective Zeeman field (see (a-d)), the nodal rings on the mutually orthogonal mirror invariant planes can be connected together to generate various shapes of nodal chains. With open boundary condition along the $z$ direction and periodic boundary condition along the $x$ and $y$ directions, the corresponding energy spectra for the systematic parameters chosen in (a-d) are shown in (e-h) respectively, where the zero-energy drumhead surface states appear at the two surfaces of $z$-direction. The corresponding topological winding numbers defined in the Brillouin zone of $z$-direction are shown in (i-l) respectively. The other parameters are  $t_{so}=t_z=t$ and $t$ is the energy unit.}
\label{Fig2}
\end{figure*}

\begin{figure*}
\centering
\includegraphics[width=16cm,height=10cm]{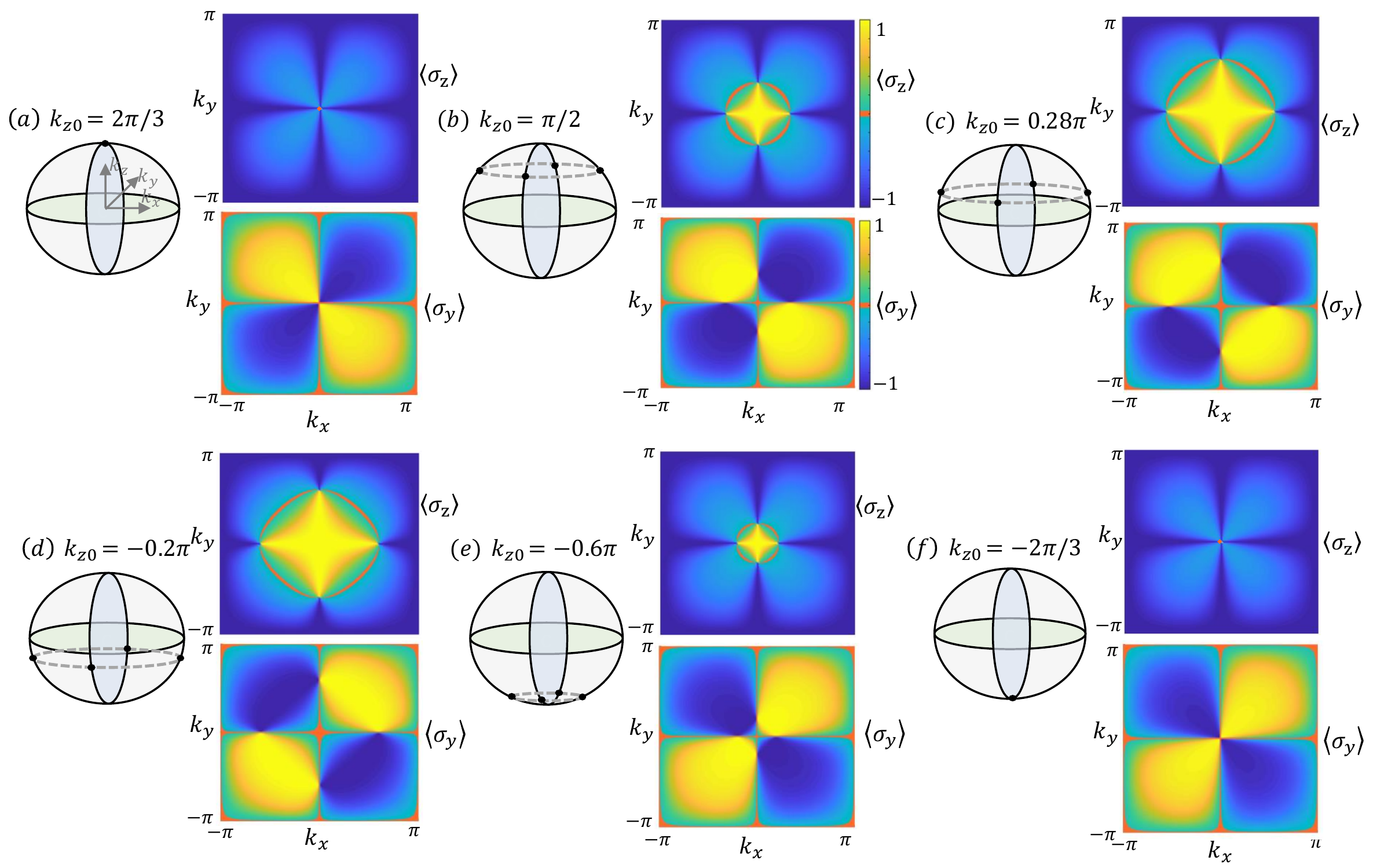}
\caption{Measuring spin polarizations $\langle\sigma_{y,z}\rangle$ to detect an inner nodal chain. The left side in (a-d) shows the shape of an inner nodal chain (formed by three connected nodal rings on the mirror invariant planes $k_{x,y,z}=0$) and the corresponding nodal points (marked by black points) when $k_z$ is fixed to $k_{z0}$. The right side in (a-d) shows the spin polarizations $\langle\sigma_{y,z}\rangle$ as a function of $k_x$ and $k_y$ for (a) $k_{z0}=2\pi/3$, (b) $k_{z0}=\pi/2$, (c) $k_{z0}=0.28\pi$, (d) $k_{z0}=-0.2\pi$, (e) $k_{z0}=-0.6\pi$ and (f) $k_{z0}=-2\pi/3$, where $\langle\sigma_{y}\rangle=0$ (marked by orange straight solid lines) and $\langle\sigma_{z}\rangle=0$ (marked by orange rings) respectively give the locations of BIS$_{1}$ and BIS$_{2}$ defined in Eqs. (\ref{BIS1}, \ref{BIS2}, \ref{BISvan}). As we can see, the nodal points for $k_z=k_{z0}$ can be detected through the intersection of the BIS$_{1}$ and BIS$_{2}$. In this way, the shape of the inner nodal chain is mapped out by measuring the vanishing of spin polarizations $\langle\sigma_{y,z}\rangle$ varying with $k_z$. The other parameters are $t_x=t_y=t_z=t_{so}=t$ and $m_z=3t$.
}
\label{Fig3}
\end{figure*}

\section{Mirror symmetry protected nodal rings and chains}
\label{secIII}

Before seeing the nodal chains emerged in the energy spectra, we transfer the real space tight-binding Hamiltonian into the momentum space to get $H(\textbf{k})=\sum_{\textbf{k}}c^{\dag}(\textbf{k})h(\textbf{k})c(\textbf{k})$,
where $c(\textbf{k})=[c_{\uparrow}(k_x,k_y,k_z),c_{\downarrow}(k_x,k_y,k_z)]^T$,
\begin{equation}
h(\textbf{k})=h_y(\textbf{k})\sigma_y+h_z(\textbf{k})\sigma_z,
\label{3DH}
\end{equation}
with
\begin{align}
&h_y=8t_{so}\sin(k_x)\sin(k_y)\sin(k_z), \nonumber \\
&h_z=m_z-2t_x\cos(k_x)-2t_y\cos(k_y)-2t_z\cos(k_z).
\label{hyz}
\end{align}
The Bloch Hamiltonian $h(\textbf{k})$ respects both the time-reversal symmetry and the inversion symmetry, i.e.,
\begin{align}
\mathcal{T}h(\textbf{k})\mathcal{T}^{-1}&=h(-\textbf{k})\nonumber \\
\mathcal{P}h(\textbf{k})\mathcal{P}^{-1}&=h(-\textbf{k})
\end{align}
where $\mathcal{T}=\mathcal{K}$ is the time-reversal symmetry operator, with $\mathcal{K}$ being the complex conjugate operator, and $\mathcal{P}=\sigma_z$ is the inversion symmetry operator. The
time-reversal symmetry requires $h_{x,z}$ being even function of $\textbf{k}$ and $h_{y}$ being odd function of $\textbf{k}$. In contrast, the inversion symmetry requires $h_{z}$ being even function of $\textbf{k}$ and $h_{x,y}$ being odd function of $\textbf{k}$. Then the consistence of these two symmetries in our system guarantees that $h_x$ vanishes in the entire Brillouin zone.

The energy spectra of $h(\textbf{k})$ is given by $E=\pm\sqrt{h^2_y+h^2_z}$ and consists of two energy bands. When $h_y(k_x,k_y,k_z)=h_z(k_x,k_y,k_z)=0$, the conduction and the valence bands could cross each other to form a chain of connected nodal rings. For our system, the nodal rings are generated on the mirror invariant planes and protected by the mirror symmetry. The reason is that the Bloch Hamiltonian $h(\textbf{k})$ also respects the mirror symmetry
\begin{align}
\mathcal{M}h(k_x,k_y,k_z)\mathcal{M}^{-1}&=h(-k_x,k_y,k_z)\nonumber \\
&=h(k_x,-k_y,k_z)\nonumber \\
&=h(k_x,k_y,-k_z),
\label{mirror}
\end{align}
where $\mathcal{M}=\sigma_z$ is the mirror symmetry operator. Note that there are six mirror invariant planes $k_{x,y,z}=0,\pi$. Now we briefly prove the existence of nodal rings on the mirror invariant planes $k_{z}=0,\pi$. The mirror symmetry in the $z$ direction in Eq. (\ref{mirror}) could lead to
\begin{align}
h_y(k_x,k_y,k_z)=-h_y(k_x,k_y,-k_z) \nonumber \\
h_z(k_x,k_y,k_z)=h_z(k_x,k_y,-k_z),
\end{align}
which indicates that $h_y(k_x,k_y,k_z)=0$ on the mirror invariant planes $k_{z}=0,\pi$. Then the solution
of $h_z(k_x,k_y,k_z)=0$ determines the two bands crossing along rings located at the mirror invariant planes $k_{z}=0,\pi$. Moreover, as we will show, these nodal rings are centered around the high symmetry points. Since the mirror operator and the Hamiltonian have same eigenstates in the mirror invariant planes, the generated nodal rings are protected by the mirror symmetry \cite{TNLrev}. Similarly, one also can prove that nodal rings also emerge on the mirror invariant planes $k_{x,y}=0,\pi$.

Specifically, for $2t<m_z<6t$, one nodal ring centered around the $\Gamma$ point is generated on the mirror invariant planes $k_{x,y,z}=0$, while there is no nodal ring generated on the mirror invariant planes $k_{x,y,z}=\pi$; For $-2t<m_z<2t$, four nodal rings centered around the $M$ point are generated on the mirror invariant planes $k_{x,y,z}=0$, and one nodal ring centered around the $X$ point is generated on the mirror invariant planes $k_{x,y,z}=\pi$; For $-6t<m_z<-2t$, one nodal ring centered around the $R$ point is generated on the mirror invariant planes $k_{x,y,z}=\pi$, while there is no nodal ring on the mirror invariant planes $k_{x,y,z}=0$. Here we assume $t_x=t_y=t_z=t_{so}=t$.

As shown in Fig. \ref{Fig2}(a-d), via tuning the laser intensities and frequencies to vary $t_x$, $t_y$ and $m_z$, the nodal rings on the mutually orthogonal mirror invariant planes could be connected together to form various nodal chains, including the outer and inner nodal chains. The outer (inner) nodal chains are generated when two nodal rings are on opposite (same) sides of the touching point \cite{NC3}. For the outer case, there are three types of outer nodal chains, including nodal nets, nodal tubes and nodal cross-lines. For example, Fig. \ref{Fig2}(a) shows that an inner nodal chain is generated by three connected nodal rings on the mutually orthogonal mirror invariant planes $k_{x,y,z}=0$. Fig. \ref{Fig2}(b) shows that an outer nodal chain shaped like a net is generated on the mutually orthogonal mirror invariant planes $k_{x,y,z}=0$ and $k_{x,y,z}=\pi$. Fig. \ref{Fig2}(c) shows that an outer nodal chain shaped like a tube is generated on the mutually orthogonal mirror invariant planes $k_{x,z}=0$ and $k_y=0,\pm\pi$, where the nodal rings on the mirror invariant planes $k_{x,z}=0$ are reduced to straight lines. Fig. \ref{Fig2}(d) shows that an outer nodal chain shaped like cross-lines is generated on the mutually orthogonal mirror invariant planes $k_{x,y}=0$ and $k_{x,y}=\pm\pi$ when $k_z=\pm\pi/2$, where all the nodal rings are reduced to straight lines.

The topological features of nodal chains are characterized by the quantized Berry phases. It is well known that the Berry phase around a single nodal line is $\pi$ \cite{TNL}. Consequently, the Berry phase around the chain point (enclosing two nodal lines) becomes 0 (=$\pi+\pi$). According to bulk-edge correspondence, such nontrivial topology implies the existence of surface states under open boundary conditions \cite{TNL}. Let us take open boundary condition along the $z$ direction and periodic boundary condition along the $x$ and $y$ directions as an example to show the generated surface states. The corresponding energy spectra is calculated in Fig. \ref{Fig2}(e-h), where we can find that the surface states appear as flat bands at the zero energy. Distinct from the drumhead surface states for nodal-ring phases, the surface states for nodal-chain phases have abundant shapes. In particular, for the nodal cross-line case shown in Fig. \ref{Fig2}(d), the regions where the surface states exist cover the whole $k_x$-$k_y$ plane.

These surface states are protected by nontrivial topological winding numbers defined in the Brillouin zone of $z$-direction, i.e.,
\begin{align}
\nu=\frac{1}{{2\pi}}\int_{-\pi}^\pi(n_y\partial_{k_z}n_z-n_z\partial_{k_z}n_y)d{k_z}
\end{align}
where $(n_y,n_z)=(h_y,h_z)/|h|$. Using the same parameters as shown in Figs. \ref{Fig2}(e-h), we calculate the values of the topological winding numbers as a function of $k_x$ and $k_y$ in Figs. \ref{Fig2}(i-l). The results show that the regions where the topological winding number values are nontrivial ($\nu=\pm1$) coincide with the regions where the surface states appear. It is worth pointing out that the sign of the nontrivial winding number values make no difference to the presence of surface states. In these regions, the corresponding Hamiltonian $H(k_x,k_y)$ supports one-dimensional inversion-symmetric topological insulators protected by a $Z_2$ topological invariant.

\section{Experimental detection of nodal chains}
\label{secIV}

The shapes of nodal chains can be detected by measuring spin polarizations. The detection principle is outlined as follows. First, on one hand, we know that the nodal chains are generated by the connected nodal rings which are determined by
\begin{equation}
h_y(k_x,k_y,k_{z})=h_z(k_x,k_y,k_{z})=0.
\label{nc}
\end{equation}
On the other hand, when $k_z$ is fixed to $k_{z0}$, the nodal chain Bloch Hamiltonian $h(k_x,k_y,k_z)$ is reduced to a two-dimensional Hamiltonian $h(k_x,k_y,k_{z0})$. In this case, there are two types of band inversion surfaces (BISs) in the $k_x$-$k_y$ plane, named as BIS$_{1}$ and BIS$_{2}$. The BISs are defined based on the vanishing of the vector fields in the Bloch Hamiltonian \cite{BIS}. Specifically,
the BIS$_{1}$ and BIS$_{2}$ are defined by
\begin{align}
h_y(k^{\text{BIS}_1}_x,k^{\text{BIS}_1}_y,k_{z0})&=0,
\label{BIS1}\\
h_z(k^{\text{BIS}_2}_x,k^{\text{BIS}_2}_y,k_{z0})&=0.
\label{BIS2}
\end{align}
Basing on Eqs. (\ref{hyz}, \ref{BIS1}, \ref{BIS2}), we can obtain the locations of BIS$_{1}$ and BIS$_{2}$. For $k_{z0}\neq0,\pi$, the BIS$_{1}$ is formed by four straight lines $k^{\text{BIS}_1}_{x,y}=0,\pi$; While for $k_{z0}=0,\pi$, the BIS$_{1}$ covers the whole $k_x$-$k_y$ plane. The BIS$_{2}$ is a ring determined by $m_z-2t_z\cos(k_{z0})=2t_x\cos(k^{\text{BIS}_2}_x)+2t_y\cos(k^{\text{BIS}_2}_y)$.
From Eqs. (\ref{nc}, \ref{BIS1}, \ref{BIS2}), we can find that, the gap-closing points for $k_z=k_{z0}$ determined by Eq. (\ref{nc}) satisfy both Eq. (\ref{BIS1}) and Eq. (\ref{BIS2}), which means that the nodal points for $k_z=k_{z0}$ are exactly located at the intersection of the BIS$_{1}$ and BIS$_{2}$. As a consequence, by scanning $k_z$ and measuring the corresponding BIS$_{1}$ and BIS$_{2}$ and their intersection, one can precisely map out the shape of the entire nodal chain.

The BISs are measured by detecting the vanishing of the spin polarizations in equilibrium. Specifically, to measure the BIS$_{1}$ and BIS$_{2}$, the spin polarizations required to be detected are
\begin{align}
\langle\sigma_s\rangle=\langle\psi_G(k_x,k_y,k_{z0})|\sigma_s
|\psi_G(k_x,k_y,k_{z0})\rangle,
\label{sp}
\end{align}
where $s=y,z$, $|\psi_G(k_x,k_y,k_{z0})\rangle$ denotes the ground state of $h(k_x,k_y,k_{z0})$. The locations of BIS$_{1}$ and BIS$_{2}$ are respectively determined by the vanishing of the spin polarizations \cite{BIS}
\begin{align}
\langle\sigma_y(k^{\text{BIS}_1}_x,k^{\text{BIS}_1}_y,k_{z0})\rangle=0,\nonumber\\
\langle\sigma_z(k^{\text{BIS}_2}_x,k^{\text{BIS}_2}_y,k_{z0})\rangle=0.
\label{BISvan}
\end{align}
The reason is that, on the BIS$_{1}$ (BIS$_{2}$), $|\psi_G\rangle$ is reduced to the ground state of $\sigma_z$ ($\sigma_y$), which exactly gives $\langle\sigma_y\rangle=0$ ($\langle\sigma_z\rangle=0$).
Therefore, for getting the BISs and their intersection varying with $k_z$, we need to perform $k_z$-resolved spin-polarization measurements in the $k_x$-$k_y$ plane.

In experiment, the time-of-flight imaging measures the $k_z$-integrated spin polarizations
\begin{equation}
\widetilde {\left\langle {{\sigma _{s}}} \right\rangle }(k_x,k_y) = \frac{1}{{2\pi }}\int_0^{2\pi } {\left\langle {{\sigma _{s}}({{k_{x}},{k_{y}},k_z} )} \right\rangle d{k_z}}.
\end{equation}
Interestingly, the recent ultracold atoms experiments in \cite{NLexp,WPexp} have demonstrated that, if the topological Hamiltonian on the BISs with respect to a $k_z$ plane (for example, $k_z=k^{c}_{z}$) respects a magnetic group symmetry, the $k_z$-resolved spin polarization $\langle\sigma_z(k^{\text{BIS}}_x,k^{\text{BIS}}_y,k^c_z)\rangle$ equals
the integral of $\langle\sigma_z(k^{\text{BIS}}_x,k^{\text{BIS}}_y,k_z)$ over $k_z$, i.e., $\widetilde {\left\langle {{\sigma _{z}}}\right\rangle}(k^{\text{BIS}}_x,k^{\text{BIS}}_y)=
\langle\sigma_z(k^{\text{BIS}}_x,k^{\text{BIS}}_y,k^c_{z})\rangle$. The experiments also developed a virtual slicing imaging technique to scan $k^c_{z}$ and measure the spin polarizations varying with $k_z$. In this way, the $k_z$-resolved spin polarizations on the BISs can be effectively detected by measuring the corresponding $k_z$-integrated spin polarizations. In the following, we will prove that the spin polarizations in our system also have such feature.

Firstly, we prove that our system also respects a magnetic group symmetry with respect to the $k_z^c=\pi/2$ plane, which ensures that the $k_z$-resolved spin polarizations $\langle\sigma_z\rangle$ on the BIS$_2$ can be effectively detected by measuring the corresponding $k_z$-integrated spin polarizations. For $k_z=\pi/2$, the BIS$_{2}$ is a ring determined by $m_z=2t_x\cos(k^{\text{BIS}_2}_x)+2t_y\cos(k^{\text{BIS}_2}_y)$. For $k_z=\pi/2+\delta k_z$, the nodal chain Hamiltonian $h(k_x,k_y,k_z)$ on this ring (BIS$_{2}$) is written as
\begin{align}
h(\delta k_z)=&8t_{so}\sin(k_{x}^{\text{BIS}_2})\sin(k_{y}^{\text{BIS}_2})\cos(\delta k_z)\sigma_y
\nonumber \\
&+2t_z\sin(\delta k_z)\sigma_z,
\end{align}
where $h(\delta k_z)=h(k^{\text{BIS}_2}_x,k^{\text{BIS}_2}_y,\pi/2+\delta k_z)$. As we can see, the nodal chain Hamiltonian respects a magnetic group symmetry with respect to the $k_z^c=\pi/2$ plane, i.e.,
\begin{equation}
\mathcal{G}h(\delta k_z)\mathcal{G}^{-1}=h(-\delta k_z),
\label{mgs}
\end{equation}
where $\mathcal{G}=\sigma_x\mathcal{K}$ is the magnetic group symmetry operator. On the other hand, we have
\begin{equation}
h(\pm\delta k_z)|\psi_{G}(\pm\delta k_z)\rangle=E_{-}|\psi_G(\pm\delta k_z)\rangle,
\label{eigfun}
\end{equation}
where $|\psi_G(\pm\delta k_z)\rangle=|\psi_G(k^{\text{BIS}_2}_x,k^{\text{BIS}_2}_y,\pi/2\pm\delta k_z)\rangle$ are the ground states of $h(\pm\delta k_z)$, and $E_{-}=E(k_z=\pi/2+\delta k_z)=E(k_z=\pi/2-\delta k_z)$. Combining Eq. (\ref{mgs}) and Eq. (\ref{eigfun}), we can obtain
\begin{equation}
|\psi_G(\delta k_z)\rangle=\mathcal{G}|\psi_G(-\delta k_z)\rangle.
\label{gs}
\end{equation}
By substituting Eq. (\ref{gs}) into Eq. (\ref{sp}), we can get
\begin{align}
\langle\sigma_z(\delta k_z)\rangle&=\langle\psi_G(\delta k_z)|\sigma_z|\psi_G(\delta k_z)\rangle
\nonumber \\
&=\langle\psi_G(-\delta k_z)|\mathcal{G}^{-1}\sigma_z\mathcal{G}|\psi_G(-\delta k_z)\rangle
\nonumber \\
&=-\langle\sigma_z(-\delta k_z)\rangle,
\end{align}
where $\langle\sigma_z(\pm\delta k_z)\rangle=\langle\sigma_z(k^{\text{BIS}_2}_x,k^{\text{BIS}_2}_y,\pi/2\pm\delta k_z)\rangle$. Then we have
\begin{equation}
 \widetilde {\left\langle {{\sigma _{z}}}\right\rangle}(k^{\text{BIS}_2}_x,k^{\text{BIS}_2}_y)=
\langle\sigma_z(k^{\text{BIS}_2}_x,k^{\text{BIS}_2}_y,k^c_z=\pi/2)\rangle.
\label{sz}
\end{equation}

Next, we prove that the mirror symmetry with respect to the mirror invariant plane $k_z=0$ guarantees that the $k_z$-resolved spin polarizations $\langle\sigma_y\rangle$ on the entire $k_x$-$k_y$ plane (not just on the BIS$_1$) have the similar feature. For $k_z=\delta k_z$, according to Eq. (\ref{mirror}), our system respects a mirror symmetry with respect to the $k^c_z=0$ plane, i.e.,
\begin{equation}
\mathcal{M}h^{\prime}(\delta k_z)\mathcal{M}^{-1}=h^{\prime}(-\delta k_z),
\label{mirror_z}
\end{equation}
where $h^{\prime}(\delta k_z)=h(k_x,k_y,\delta k_z)$. Similarly, we can prove that
\begin{align}
|\psi^{\prime}_G(\delta k_z)\rangle&=\mathcal{M}|\psi^{\prime}_G(-\delta k_z)\rangle,\nonumber \\
\langle\sigma_y(\delta k_z)\rangle&=-\langle\sigma_y(-\delta k_z)\rangle,
\end{align}
where $\langle\sigma_y(\pm\delta k_z)\rangle=\langle\sigma_y(k_x,k_y,\pm\delta k_z)\rangle$. Then we get
\begin{equation}
 \widetilde {\left\langle {{\sigma _{y}}}\right\rangle}(k_x,k_y)=
\langle\sigma_y(k_x,k_y,k^c_{z}=0)\rangle.
\label{sy}
\end{equation}
Therefore, basing on Eqs. (\ref{sz}, \ref{sy}) and the virtual slicing imaging technique \cite{NLexp,WPexp}, the $k_z$-resolved spin polarizations on the BISs could be effectively measured. From this measurement, we can acquire the BISs and their intersection and map out the shapes of nodal chains.

As a concrete example, Fig. \ref{Fig3} presents how the shape of the inner nodal chain (see Fig. \ref{Fig2}(a)) is detected by measuring $k_z$-resolved spin polarizations. The corresponding spin polarizations $\langle\sigma_{y,z}\rangle$ versus $k_x$ and $k_y$ for different $k_{z0}$ are numerically calculated in Figs. \ref{Fig3}(a-f). For $k_{z0}=2\pi/3$, the results in Fig. \ref{Fig3}(a) show that $\langle\sigma_{y}\rangle=0$ when $k_{x,y}=0,\pi$ and
$\langle\sigma_{z}\rangle=0$ when $k_x=k_y=0$. Then we can find that the BIS$_{1}$ locates at $k^{\text{BIS}_1}_{x,y}=0,\pi$ and the BIS$_{2}$ at $k^{\text{BIS}_2}_x=k^{\text{BIS}_2}_y=0$.
Through the intersection of the BIS$_{1}$ and BIS$_{2}$, we can detect the nodal point as ($k_x,k_y,k_z$)=($0,0,2\pi/3$). For $k_{z0}=\pi/2$, the results in Fig. \ref{Fig3}(b) show that the BIS$_{1}$ still locates at $k^{\text{BIS}_1}_{x,y}=0,\pi$, while the BIS$_{2}$ locates on a ring determined by $\cos(k^{\text{BIS}_2}_x)+\cos(k^{\text{BIS}_2}_y)=3/2$. The intersection of the BIS$_{1}$ and BIS$_{2}$ allows us to detect the four nodal points as ($k_x,k_y,k_z$)=($0,\pm\pi/3,\pi/2$),($\pm\pi/3,0,\pi/2$). The results in Figs. \ref{Fig3}(c-f) also demonstrate this point, i.e., the nodal points associated with different $k_{z0}$ can be directly mapped out by measuring the $k_z$-resolved spin polarizations and extracting the intersection of the BIS$_{1}$ and BIS$_{2}$. In this way, the shape of the inner nodal chain is detected. This method is general and also can be applied to detect the shapes of the outer nodal chains illustrated in Fig. \ref{Fig2}(b-d).
\\
\\

\section{Summary}
\label{secV}

In summary, we have presented an experimental scheme to realize topological nodal chain semimetal phases with ultracold atoms trapped in optical Raman lattices. In our scheme, we have constructed a three-dimensional optical Raman lattice that could produce next nearest-neighbor spin-orbit couplings and host various shapes of mirror symmetry protected nodal chains in its energy spectra. Finally, we have shown that the shapes of the nodal chains could be detected by measuring spin polarizations. Thus, our study has demonstrated the potential of optical Raman lattice systems as a versatile platform for exploring topological nodal-chain semimetal phases.

Moreover, our study sets an example for realizing complex spin-orbit couplings in optical Raman lattices by applying suitable Raman potentials. This strategy is expected to overcome the challenges faced in generating the complex spin-orbit couplings required for realizing high-dimensional topological phases in optical lattice systems. A straightforward example is that the topological nodal links and knots \cite{NC8} could be realized by adding perturbations into the Raman potentials applied in our nodal chain platform. In the future, it would be quite interesting to apply this strategy in optical Raman lattices to construct the spin-orbit couplings for realizing the four-band topological phases, including the $Z_2$ topological insulator phases \cite{Z21,Z22}, topological Dirac semimetal phases \cite{DS1,DS2} and higher-order topological phases \cite{HOT}.

\section{Acknowledgment}

B.B.W. and J.H.Z. contributed equally to this work. This work was supported by the National Key Research and Development Program of China (2017YFA0304203), National Natural Science Foundation of China (NSFC)
(12034012, 12074234), Research Program of the Education Department of Hubei province (D20181903), Changjiang Scholars and Innovative Research Team in University of Ministry of Education of China (PCSIRT)(IRT\_17R70), Fund for Shanxi 1331 Project Key Subjects Construction, and 111 Project (D18001).

\end{document}